\documentclass[%
 reprint,
 amsmath,amssymb,
 aps,
]{revtex4-1}
\usepackage{graphicx}
\usepackage{dcolumn}
\usepackage{bm}

\begin{document}


\title{Effective Control of Cold Collisions with Radio Frequency Fields}

\author{Yijue Ding$^1$, Jos\'e P. D'Incao$^{2,3}$ and Chris H. Greene$^1$}
\email[]{ding51@purdue.edu}
\affiliation{$^1$Department of Physics and Astronomy, Purdue University, West Lafayette, Indiana, USA \\
$^2$ JILA, University of Colorado and NIST, Boulder, Colorado, USA\\
$^3$ Department of Physics, University of Colorado, Boulder, Colorado, USA}


\date{\today}

\begin{abstract}
We study $^{87}$Rb cold collisions in a static magnetic field and a single-color radio frequency (RF) field by employing the multi-channel quantum defect theory in combination with the Floquet method to solve the two-body time-dependent Schr\"odinger equation. Our results show that RF fields can modify the two-body scattering length by a large scale through Feshbach resonances both in low and high static magnetic field regimes. Such RF induced Feshbach resonances can be applied to quenching experiments or controlling interactions in spinor condensates. Here, we also show that analogous to photo-association, RF fields can also associate cold atoms into molecules at a useful rate. 
\end{abstract}


\maketitle
Ultracold quantum gases are simple and clean systems that we can control easily and precisely. Recently various techniques such as synthetic fields, optical lattices and cavities have been developed in order to achieve novel quantum regimes\cite{synfieldreview,optlatticenatphys,cavityrmp}. At the two-body level, the use of Feshbach resonances is the most powerful tool to control interactions of quantum gases, characterized by the $s$-wave scattering length\cite{juliennermp,chengchinrmp}. The magnetic Feshbach resonance (MFR) is commonly used in alkali metal species, where atomic Zeeman levels are tuned to make the bound state and the scattering state nearly energy degenerate\cite{moerdijkmfr}. Laser fields are also employed to create resonances in systems without intrinsic channel coupling\cite{junyeofr, killianbec} or as an auxiliary to MFR\cite{theoryofr, heckerofr, takahashiofr}, which is  called the optical Feshbach resonance (OFR). However, the application of OFR is hampered by large atom loss, heating and small tunability. Besides, there are still many systems such as alkaline earth and transient atoms that limit our ability to explore. Some new ideas such as orbital interaction induced Feshbach resonances have been introduced to control interactions in these systems\cite{orbitalfrtheo,orbitalfrexp}.

Our major purpose in this article is to propose radio frequency (RF) fields as an effective tool for controlling cold collisions in systems where the MFR or OFR are not useful or even not attainable. RF has been used in the cold atom realm for several purposes such as Efimov trimer spectroscopy\cite{Lomperfefimovexp,rfefimovtheory,ueda2011} and RF dressed state trapping\cite{rftrap}. One significant advantage of RF is that it is easy to control and manipulate compared to laser fields or magnetic fields. Therefore, it can be applied to spinor condensates\cite{stamperrmp} where the magnetic field is nearly vanishing or improve quenching experiments\cite{makotynquench} with a faster ramping speed than magnetic fields. It can also apply to alkaline earth species that have negligible intrinsic channel coupling. 

Here we discuss $^{87}$Rb collisions in an external magnetic field and a single-color RF field. The Hamiltonian representing the relative motion of two identical $^{87}$Rb atoms is given by
\begin{equation}
H=-\frac{\hbar^2}{2\mu}\nabla^2+\hat{V}+H_{Zeeman}+H_{hf}+W(t),
\end{equation}
where $\mu$ is the two-body reduced mass, $\hat{V}$ is the spin dependent interatomic potential and $W(t)$ denotes the interaction between the atom and the RF fields. Since the RF field is homogeneous in space and its duration is sufficiently long, this interaction $W(t)$ is considered to be space independent and has a temporal period $T=1/\nu$, 
where $\nu$ is the period of the RF field. This periodicity allows us to use Floquet theory to treat this time-dependent Hamiltonian\cite{chu20041,floquetbook}, which facilitates the problem by converting the time-dependent Schr\"odinger equation to a time-independent one. We solve the resulting coupled channel equations using the multi-channel quantum defect approach\cite{greenermp,seatonqdt}, which has proved its robustness in cold collisions\cite{ruzicmqdt,bogaomqdt}. Our model is further simplified with the frame transformation approximation\cite{burkemqdt}. The RF frequency ranges from MHz to GHz, which drives transition between atomic hyperfine levels or Zeeman sublevels.  For a $\sigma_x$ polarized RF field, the interaction is written in terms of the two atomic magnetic moment operators $\vec{\mu}_i(i=1,2)$ and magnetic amplitude $\vec{B}_{RF}=B_{RF}\hat{x}$:
\begin{equation}
W(t) =-\left(\vec{\mu}_1\cdot \vec{B}_{RF}+\vec{\mu}_2\cdot \vec{B}_{RF}\right)\text{cos}\left(2\pi\nu t\right).
\end{equation}
Moreover, the selection rule of RF induced transitions is $\Delta n=\pm 1, \Delta m_f=\pm 1$ , where $n$ is the total photon number in the space(ambient photon number). This indicates that one atom will flip spin by absorbing or emitting a photon. 

 The RF photon can resonantly interact with atoms through three major processes: a free-free transition, a bound-free transition or a bound-bound transitions, as illustrated in Fig. 1 and indicated as cases (i), (ii) and (iii), respectively. In each case, the strength of the RF induced transition will depend on the RF amplitude ($B_{RF}$), the Franck-Condon overlap between the relevant atomic and molecular states, and the detuning. Here we will explore the control of atomic collisions via bound-bound and bound-free transitions, since those are the cases in which one can expect the formation of field-induced Feshbach resonances.

In the RF induced bound-bound transition scenario, RF photons drive transitions between two weakly bound molecules. The bare Rabi frequency between two weakly bound molecules, which characterizes the strength of the transition in zero detuning, is given by 
\begin{equation}
\hbar\Omega= | \langle\psi_{b1}|\vec{\mu} \cdot \vec{B}_{RF}|\psi_{b2}\rangle |,
\end{equation}
where $\vec{\mu}=(\vec{\mu}_1+\vec{\mu}_2)$ is the total magnetic moment of two free atoms, $\psi_{b1}$ and $\psi_{b2}$  are bound state wave functions including all spin and spatial degrees of freedom.

A good candidate to explore this scenario occurs for $^{87}$Rb near B=1008G. Since there is a pre-existing Feshbach resonance located at B=1008.8G~\cite{rb87fr} originating in the interaction between the $|1,1\rangle+|1,1\rangle$ scattering state and a bound state (20,22) [we label the molecular states by $(f_1 m_{f1},f_2 m_{f2})$ and omit the vibrational quantum number], that scattering state is first magneto-associated into the weakly bound state (20,22), then this molecule can couple to another molecular state (2-1,22) by coherently emitting a RF photon near $\nu=$745MHz. This coupling produces two dressed molecular states separated by the Rabi frequency, forming an Autler-Townes splitting feature \cite{autlertownes} in the resonance profile as shown in Fig. 2(a). A similar Autler-Townes doublet feature was also observed in laser controlled resonances~\cite{laserdoublet}, though strong atomic losses were observed due to spontaneous decay. Figure 2(b) shows the real part of the scattering length as a function of the magnetic field at $\nu=745$MHz, which also shows the Autler-Towns splitting of the pre-existing resonance. In our case, however, since both of the bound states are below the collision threshold for fields near the Feshbach resonance position, there are no inelastic collisions due to spin relaxation. Moreover, because the RF frequency is far detuned from any other channel thresholds and the Franck-Condon factors between those molecular states and the inelastic scattering states are fairly small, the RF dissociation to inelastic channels is negligible. Therefore, elastic scatterings dominate the collision process and the scattering length is almost divergent near the RF-induced resonances. 
On the other hand, scanning the RF frequency at a constant static magnetic field B=1009.2G yields a Feshbach resonance profile(not shown) as a function of the RF frequency with a resonant position $\nu_0\approx744.4$MHz and a corresponding frequency-width $\Delta\nu\approx0.4$MHz. Such resonances can be used in quenching experiments \cite{makotynquench} which can help to overcome the difficulties of changing magnetic fields rapidly. 

\begin{figure}
\label{fig1}
\includegraphics[width=0.45\textwidth]{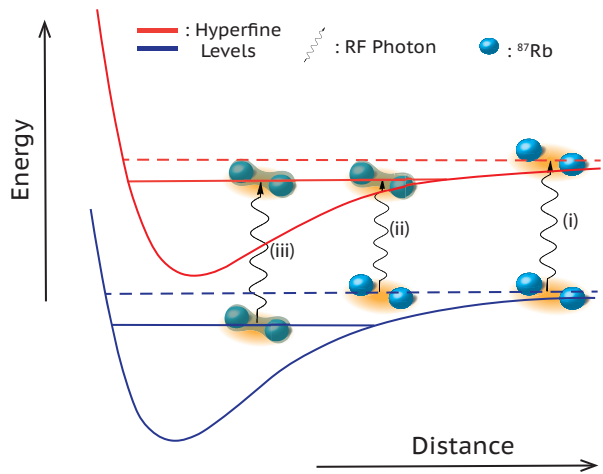}
\caption{Resonant interactions between RF fields and cold atoms. In ultra-cold regime, collisions are always near channel thresholds as denoted by the horizontal dashed lines. The horizontal solid lines represent the weakly bound states of each channel. There are three major processes of RF induced transition between $^{87}\text{Rb}$ hyperfine levels:(i) free-free transitions in the asymptotic region, (ii) bound-free transitions in the Franck-Condon overlapping area, and (iii) bound-bound transitions in the Franck-Condon overlapping area and also at short distance.}

\end{figure} 
\begin{figure}
\label{fig2}
\includegraphics[width=0.48\textwidth]{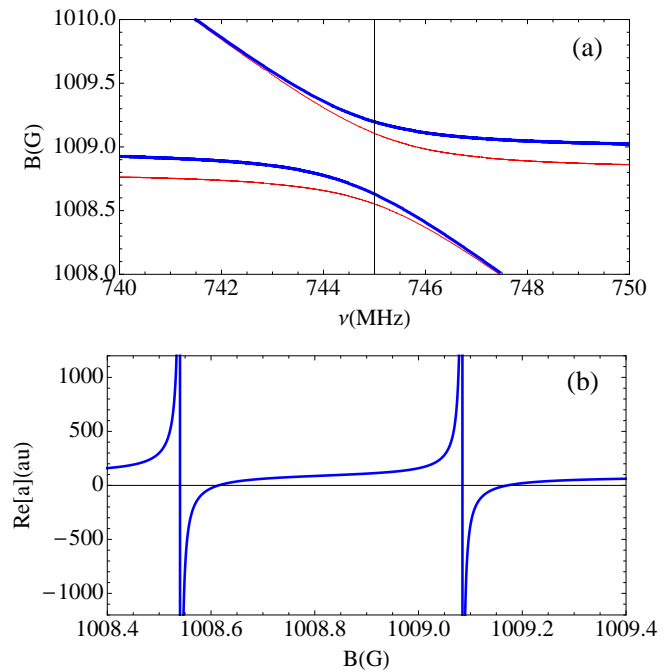}
\caption{(a) Resonance profile as a function of the magnetic field and the frequency of the RF field with amplitude $B_{RF}=5$G for scattering channel $|1,1\rangle+|1,1\rangle$ at 1$\mu$K. Thin red line: positions where the scattering length is nearly divergent. Thick blue line: positions where the scattering length vanishes. The avoided crossing indicates a RF induced transition between two bound states. (b) The real part of the scattering length as a function of the magnetic field at $\nu=745$MHz, as marked by the vertical line in (a).}
\end{figure} 

In the absence of any pre-existing Feshbach resonances, RF fields can also effectively control the scattering lengths via RF induced bound-free transitions. This approach is analogous to the one used for optical Feshbach resonances \cite{junyeofr,nicholsonofr} except here there will be negligible spontaneous emission. We explore this process for $^{87}$Rb at small magnetic fields $B\le30$G to enable the creation of strongly interacting spinor condensates\cite{colussispinorbec,colussispinor2}. The choice of the molecular state is crucial and we seek molecular states with reasonable Franck-Condon overlap with incident channels. Promising molecular states appear for RF frequencies around 6.8GHz, where transitions occur between atomic and molecular states in the $f=1$ and $f=2$ hyperfine states of $^{87}$Rb. Now, since those molecular states can dissociate, the tunability of the scattering length is limited by the onset of this inelastic process. 
For studies of spinor condensates, however, since the ferromagnetic and anti-ferromagnetic regimes are very sensitive to the values of the scattering lengths\cite{stamperrmp}, even a small change of the bare scattering length can lead to a substantial modification of the underlying many-body phenomena, provided that the time scale for the atomic losses is not too drastic.

In the presence of inelastic processes, the scattering length acquires an imaginary part while the real part characterizes the two-body interaction strength in the low energy limit\cite{dalgarno1997}. For near threshold collisions where $ka_{bg}\ll1$, the maximum tunability of the scattering length $\Delta a$ is given by \cite{hutsonnjp, scatteringtheory}
\begin{equation}
\label{deltaa}
\Delta a=\text{Re}[a]_{max}-\text{Re}[a]_{min}=\Gamma_{incid}/(k\Gamma_{in}),
\end{equation}
where $\Gamma_{incid}$ is the partial width of the incident channel,
$\Gamma_{in}=\sum_{j\neq incid}\Gamma_j $ is the total inelastic width and  $\Gamma_j$ is the partial width for each inelastic channel. The molecular lifetime is simply given by $\tau=1/(\Gamma_{incid}+\Gamma_{in})$. 
Thus, a molecular state with strong coupling to the incident channel while having a weak coupling to inelastic channels is preferable to induce resonances and control the scattering process. Previous work in Refs. \cite{rfinduce,kaufmanrf,dalibardmwfr} has found small tunability of scattering lengths within the chosen range of
parameters and the systems studied, which has limited its impact on experiments. 

In the presence of inelastic processes, the scattering length dependence on the RF frequency can be described by
\begin{equation}
\label{vdepfinite}
a=a_{bg}\left(1-\frac{\Delta\nu}{\nu-\nu_0-i\Gamma_{in}/2}\right).
\end{equation}
As a result, the resonance profile can be completely characterized by four parameters $a_{bg}$, $\nu_0$, $\Delta\nu$ and $\Gamma_{in}$. 

Table I exhibits examples of RF induced resonances with relatively large tunability of scattering lengths from different incident channels for magnetic fields ranging from $B=0.2$ to $30$G and for RF amplitudes $B_{RF}=5$ or $10$G. These resonances can be employed to effectively control two-body interactions in cold gases. The results listed in Table I imply molecular lifetimes ($\tau=1/(\Gamma_{incid}+\Gamma_{in})$) about 2ms to 60ms, largely exceeding typical molecular lifetimes (1ns-1$\mu$s) used for optical Feshbach resonances.
\begin{table*}
\label{rfinducedrf}
\caption{RF induced Feshbach resonances of $^{87}$Rb showing large scattering length tunability for various magnetic field strengths and RF amplitudes at 1$\mu K$. These examples correspond to RF induced transitions between a continuum state and a molecular state in different hyperfine levels. The list shows the incident channel indices, the external magnetic field strength, the RF amplitude, the tunability of scattering lengths and resonance parameters in Eq.(\ref{vdepfinite}).}
\begin{center}
\begin{ruledtabular}
\begin{tabular}{cccccccc}
channel &  B(G) & $B_{RF}$(G) & $\Delta a$(au) & $\nu_0$(MHz) & $\Delta \nu$(Hz) & $\Gamma_{in}$(Hz) &$\Gamma_{incid}$(Hz) \\ \hline
$|1,1\rangle+|1,1\rangle$   &0.2 & 5 & 164 & 6810.334 & 44 & 51 & 5.9\\
$|1,1\rangle+|1,1\rangle$   & 6 & 10 & 75 & 6823.158 & 170 & 420 & 22.3\\
$|1,1\rangle+|1,1\rangle$   & 10 & 5 & 27 & 6830.963 & 41 & 277 & 5.3\\
$|1,-1\rangle+|1,-1\rangle$   & 1 & 5 & 225 & 6807.805 & 42 & 34 &5.4\\
$|1,-1\rangle+|1,-1\rangle$   & 6 & 5 & 362 & 6797.303 & 43 & 22 &5.6\\
$|1,-1\rangle+|1,-1\rangle$    & 20 & 10 & 950 & 6768.528 & 172 & 34 &22.8\\
$|1,-1\rangle+|1,-1\rangle$    & 30 & 10 & 1542 & 6747.614 & 174 & 21 &22.9\\
$|1,-1\rangle+|2,-2\rangle$   & 0.2 & 5 & 465 & 6859.580 & 42 & 17 &5.6\\
$|1,-1\rangle+|2,-2\rangle$   & 1 & 5 & 336 & 6857.901 & 42 & 23 &5.5\\
$|1,-1\rangle+|2,-2\rangle$   & 6 & 10 & 128 & 6846.761 & 164 & 240 &21.7\\
\end{tabular}
\end{ruledtabular}
\end{center}
\end{table*}

\begin{figure}
\label{amplitudedep}
\includegraphics[width=0.48\textwidth]{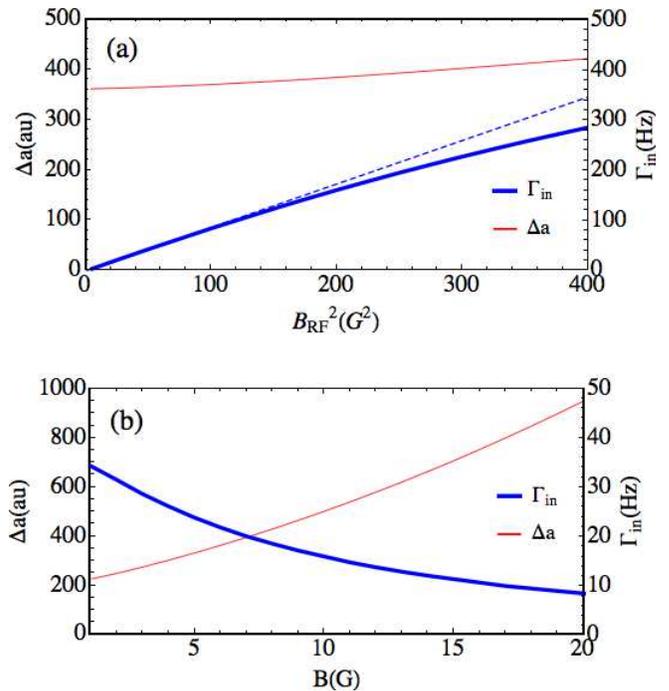}
\caption{Tuning Feshbach resonances with the RF amplitude and the static magnetic field strength. All collisions are from $|1,-1\rangle+|1,-1\rangle$ channel at 1$\mu$K. (a) shows the tunability of scattering lengths $\Delta a$ (thin red), the inelastic width $\Gamma_{in}$ (thick blue) as a function of RF amplitude square in a static magnetic field $B=6$G. The blue dashed line indicates a linear fitting of $\Gamma_{in}$ at small RF amplitude. (b) shows the tunability of scattering lengths and the inelastic width as a function of the magnetic field strength with a fixed RF amplitude $B_{RF}=5$G.}
\end{figure}

As an example, we single out our study for collisions between atoms in the $|1,-1\rangle$ hyperfine state at $E=1\mu$K with a RF field frequency tuned near 6.8GHz. 
A few interesting features are obtained as the RF amplitude varies. 
First, the resonance position $\nu_0$ varies quadratically with the RF B-field amplitude, which is analogous to the AC stark effect in optical photo association. Second, since both the partial width of incident channel $\Gamma_{incid}$  and the total inelastic width $\Gamma_{in}$ vary quadratically with the RF amplitude (according to the Fermi golden rule), one cannot effectively change the tunability of scattering lengths by increasing the RF intensity. As shown in Fig. 3(a), $\Delta a$ only increases by 15\% when the RF amplitude varies from 2G to 20G. 
In the low intensity region, the inelastic width varies almost linearly with the RF intensity (proportional to $B_{RF}^2$), which indicates that the RF dissociation prevails over the spin relaxation and dominates the inelastic process. At higher RF intensities, the inelastic width deviates from linearity. This is because the association saturates at very high intensity, as known in the optical photo association~\cite{pasaturation}. But within the intensity range of Fig. 3(a), the inelastic width keeps increasing, showing that the intensity is not high enough to saturate the RF association process. 

Eq. (\ref{deltaa}) shows that a smaller inelastic width (i.e., longer molecular lifetime) leads to larger tunability of scattering lengths. In fact, an improvement of molecular lifetime can be obtained by increasing the static magnetic field strength, as shown in Fig. 3(b). At B=20G, the tunability of scattering lengths reaches as large as 1000 a.u., corresponding to a molecular lifetime of about 100ms. 
With increasing the static magnetic field, the Zeeman level spacing becomes larger and thereby the molecular state dissociates into inelastic channels with larger kinetic energies, which is a less probable process due to 
the reduction of the Franck-Condon overlap between the molecular state and the inelastic channel wave functions. 
On the other hand, since collisions are always near threshold, the Franck-Condon factor between the molecular state and the incident channel wave functions 
keeps large. 
Consequently, the total inelastic width drops while the tunability of scattering lengths increases significantly with the increment of the static magnetic field. 

Due to the extended molecular lifetimes in RF induced resonances we also explore the production of ultra-cold molecules through RF association processes\cite{dsjinrfasso}. It is known that forming homonuclear molecules through optical photo association (PA) is restricted by its low efficiency because of the small Franck-Condon factor and large spontaneous decay\cite{pareview}. These difficulties can, however, be avoided using RF fields. Adapting standard PA theory but neglecting spontaneous emissions, the RF association rate of two identical bosons is given by 
\begin{equation}
K_2=\frac{\pi g^{(2)} \hbar}{\mu k}\frac{\Gamma_{incid}\Gamma_{in}}{(\nu-\nu_0)^2+(\Gamma_{incid}+\Gamma_{in})^2/4},
\end{equation}
where the coefficient $g^{(2)}=2$ for thermal cloud of identical bosons and $g^{(2)}=1$ for BEC\cite{chengchinrmp}.
\begin{figure}
\includegraphics[width=0.48\textwidth]{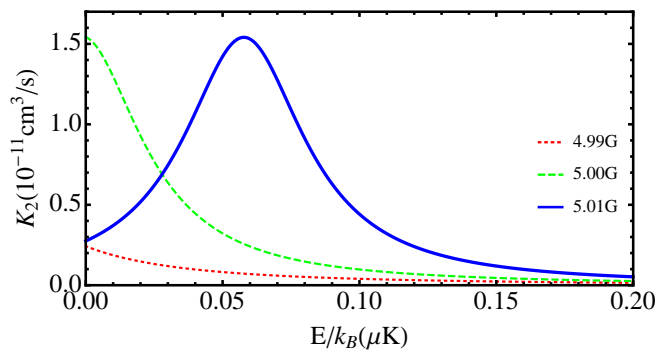}
\caption{The RF association rate for collisions between $|1,1\rangle$ states in BEC regime as a function of the collision energy for various RF amplitudes $B_{RF}=$4.99G(red dotted), 5.00G(green dashed) and 5.01G(blue solid) at $\nu=6841.59$MHz and $B=15$G.}
\end{figure}

Table II shows the RF association parameters for various collision channels at $B=6$G. The maximum rate coefficient for some channels can reach $10^{-10}\text{cm}^3/\text{s}$, much larger than the typical PA rate ($10^{-12}\text{cm}^3/\text{s}$) in a $^{87}$Rb spinor condensate\cite{rb87spinorbec}. Therefore, RF fields can be an efficient method of producing ultra-cold molecules. Since partial width of the incident channel $\Gamma_{incid}$ is proportional to the incident momentum $k$ at small energies, the Wigner threshold law yields a constant value for the rate coefficient at vanishing collision energies as shown in Fig. 4. Again it demonstrates the shift of resonance profiles as a function of RF amplitudes.

\begin{table}
\label{rfastable}
\caption{RF association profiles for $f=1$ states at magnetic field $B$=6G and with the RF amplitude $B_{RF}$=5G. The collision energy is 1$\mu$K for all cases.}
\begin{center}
\begin{ruledtabular}
\begin{tabular}{ccccc}
channel & $\nu_0$(MHz) &$K_2^{max}$($\text{cm}^3/\text{s}$) &$\Gamma_{in}$(kHz) &$\Gamma_{incid}$(Hz)\\
\hline
$|1,1\rangle$+$|1,1\rangle$ & 6822.54 & $1.39\times 10^{-10}$  & 0.1 &5.7\\
$|1,1\rangle$+$|1,0\rangle$ & 6823.12 & $1.04 \times 10^{-12}$ & 21 &7.9\\
$|1,0\rangle$+$|1,0\rangle$ & 6813.90 & $1.66\times 10^{-12} $ & 4.6 &2.8\\
$|1,-1\rangle$+$|1,-1\rangle$ & 6797.30 & $4.51 \times 10^{-10}$ & 0.02 &5.1\\
\end{tabular}
\end{ruledtabular}
\end{center}
\end{table}

In summary, we have developed a theoretical treatment of radio frequency assisted cold collisions for $^{87}$Rb. The RF fields can manipulate existing Feshbach resonances and create new resonances. The RF induced bound-bound transition modifies the pre-existing resonance profile near $B=1008$G, forming an Autler-Townes splitting feature. In the low magnetic field region, specifically $B\le30$G, a few candidates out of abundant RF induced Feshbach resonances yield large tunability of scattering lengths. Moreover, the RF field is an effective tool of forming ultra-cold molecules in terms of high association rates. Based on our results, the RF field may be an innovative and promising approach to controlling cold atom interactions. 

CHG thanks Dan Stamper-Kurn for discussions that sparked the present study.  This work was supported in part by NSF Grant PHY-1306905, NSF
Grant PHYS-1307380 and by  BSF Grant 2012504
\bibliography{apstemplate.bib}

\end{document}